\documentclass[aps,prc,twocolumn,superscriptaddress,nofootinbib,longbibliography]{revtex4-2}

\usepackage[T1]{fontenc}
\usepackage[utf8]{inputenc}
\usepackage{graphicx}
\usepackage{amsmath,amssymb,bm}
\usepackage{booktabs}
\usepackage{array}
\usepackage{tabularx}
\usepackage{siunitx}
\usepackage[colorlinks=true,linkcolor=blue,citecolor=blue,urlcolor=blue]{hyperref}

\newcommand{\mucf}{\ensuremath{\mu\mathrm{CF}}}
\newcommand{\dtmu}{\ensuremath{dt\mu}}

\usepackage{booktabs}
\usepackage{makecell}

\begin{document}
	
	\title{Muon-Catalyzed Nuclear Fusion: Physical Mechanism, Bottleneck Breakthroughs, and an Engineering Pathway}
	
	\author{Xiong Yin}
	\email{yxsxlyzh@163.com}
	\affiliation{
		Shenzhen Haihean Science Research Co., Ltd.,
		Shenzhen 518031, Guangdong, China
	}
	
	\author{Wei Kou}
	\affiliation{
		State Key Laboratory of Heavy Ion Science and Technology,
		Institute of Modern Physics, Chinese Academy of Sciences,
		Lanzhou 730000, China
	}
	\affiliation{
		School of Nuclear Science and Technology,
		University of Chinese Academy of Sciences,
		Beijing 100049, China
	}
	\affiliation{
		Southern Center for Nuclear Science Theory,
		Institute of Modern Physics, Chinese Academy of Sciences,
		Huizhou 516000, China
	}
	\author{Xurong Chen}
	\email[Corresponding author: ]{xchen@impcas.ac.cn}
	\affiliation{
		State Key Laboratory of Heavy Ion Science and Technology,
		Institute of Modern Physics, Chinese Academy of Sciences,
		Lanzhou 730000, China
	}
	\affiliation{
		School of Nuclear Science and Technology,
		University of Chinese Academy of Sciences,
		Beijing 100049, China
	}
	\affiliation{
		Southern Center for Nuclear Science Theory,
		Institute of Modern Physics, Chinese Academy of Sciences,
		Huizhou 516000, China
	}

	\begin{abstract}
		Muon-catalyzed nuclear fusion (\mucf) replaces atomic electrons with negative muons, compressing atomic orbitals by about two orders of magnitude and enabling deuterium--tritium (D--T) fusion under near-room-temperature conditions.
		This paper reviews the physical principles of \mucf{} and formulates its essential dynamics as a four-step cycle: muonic-atom formation, muon transfer, resonant \dtmu{} molecular formation, and D--T fusion with muon release and recycling.
		A kinetic model is used to quantify the number of catalysis cycles per muon and the corresponding energy gain.
		We focus on the central limitation of catalytic efficiency, namely the alpha-sticking effect, and discuss possible breakthrough routes including nuclear-spin and muon dual polarization, in-flight muon-catalyzed fusion, and heavy-ion-driven magneto-inertial fusion.
		Within the idealized assumptions of the present model, a four-dimensional synergistic scheme combining dual polarization, high-density confinement, electric-field-assisted muon recovery, and resonant enhancement may increase the number of catalysis cycles per muon from the present experimental record of about 150 to more than 500, potentially enabling an energy gain \(Q>2\).
		On this basis, we propose a conceptual fusion--fission fuel-breeding hybrid reactor, denoted as \mucf-FBR, which exploits the 14.1-MeV neutron yield of \mucf{} to breed \({}^{239}\mathrm{Pu}\) from a \({}^{238}\mathrm{U}\) blanket in a decoupled fusion--fission operating mode.
		This concept may offer advantages in engineering robustness, radiation-damage tolerance, and natural-uranium utilization.
	\end{abstract}
	
	
	\maketitle
	
	\section{Introduction}
	
	Achieving controlled nuclear fusion has long been regarded as a ``holy grail'' for solving the energy problem of humanity.
	The dominant routes in present global fusion research are magnetic-confinement fusion, exemplified by tokamak devices, and inertial-confinement fusion, exemplified by laser-ignition facilities.
	Both approaches essentially use large-scale engineering devices to reproduce the extremely high-temperature and high-pressure environment inside stellar cores, thereby overcoming the Coulomb repulsion between atomic nuclei \cite{Wesson2011}.
	However, both routes face severe physical and engineering challenges.
	Magnetic-confinement fusion requires the confinement of plasma at temperatures of order \(10^{8}\) K and the realization of high-parameter steady-state operation, whereas inertial-confinement fusion is limited by laser-conversion efficiency, target-fabrication precision, and low-repetition-rate operation \cite{Hurricane2014}.
	In essence, both routes attempt to reconstruct, on the laboratory scale, the gravitational-confinement conditions of stellar cores.
	They therefore inevitably encounter systematic challenges such as large device size, construction costs above the ten-billion-dollar scale, and commercialization time scales of several decades. Despite its elegant microscopic mechanism, \mucf{} has not yet become a practical energy technology because the finite muon lifetime, the high energy cost of muon production, and muon loss through alpha sticking strongly limit the achievable net energy gain.
	These limitations motivate the kinetic and engineering analyses developed below.
	
	Beyond these two ``hundred-million-degree'' routes, one may ask whether a ``room-temperature route'' exists.
	Muon-catalyzed nuclear fusion gives an affirmative answer.
	The core idea is profound yet simple: if the Sun relies on gravity to compress matter, can a fundamental particle much heavier than the electron play an analogous role on the microscopic scale by compressing atomic orbitals?
	In 1947, Frank published a pioneering paper in \textit{Nature}, demonstrating from quantum-mechanical arguments that negative muons could reduce the D--T internuclear distance through orbital compression and thereby bring the nuclei into the range where fusion becomes possible \cite{Frank1947}.
	This work laid the theoretical foundation of \mucf{}, and the idea was subsequently developed by Sakharov and Zel'dovich \cite{Sakharov1948,Zeldovich1954}.
	In brief, when a negative muon with a mass about 207 times that of the electron is injected into a D--T mixture, it replaces an electron and forms a tightly bound muonic atom.
	Because the orbital radius scales inversely with the orbiting-particle mass, the internuclear distance in a \(\dtmu\) molecule is compressed to approximately \(280~\mathrm{fm}\), only about \(1/264\) of the internuclear distance of an ordinary hydrogen molecule, approximately \(74000~\mathrm{fm}\).
	As a result, the quantum tunneling probability between the deuteron and triton is dramatically enhanced, and fusion can occur spontaneously under near-room-temperature conditions \cite{Jackson1957}.
	
	This paper systematically discusses the physical mechanism and key kinetic processes of \mucf{} and reviews the research history over the past seven decades.
	By constructing a quantitative model of the catalytic cycle, we focus on frontier technical schemes for overcoming the present energy-balance bottleneck.
	Finally, we analyze the \mucf-FBR fusion--fission hybrid concept proposed here and discuss its engineering prospects.
	The phrase ``miniature Sun'' is used only as a heuristic metaphor for muon-induced microscopic compression, not as an assertion that \mucf{} reproduces stellar thermodynamic conditions.
	
	\section{Physical Mechanism of Muon Catalysis: A Four-Step Cycle}
	
\begin{figure*}[t]
		\centering
		\includegraphics[width=0.498\textwidth]{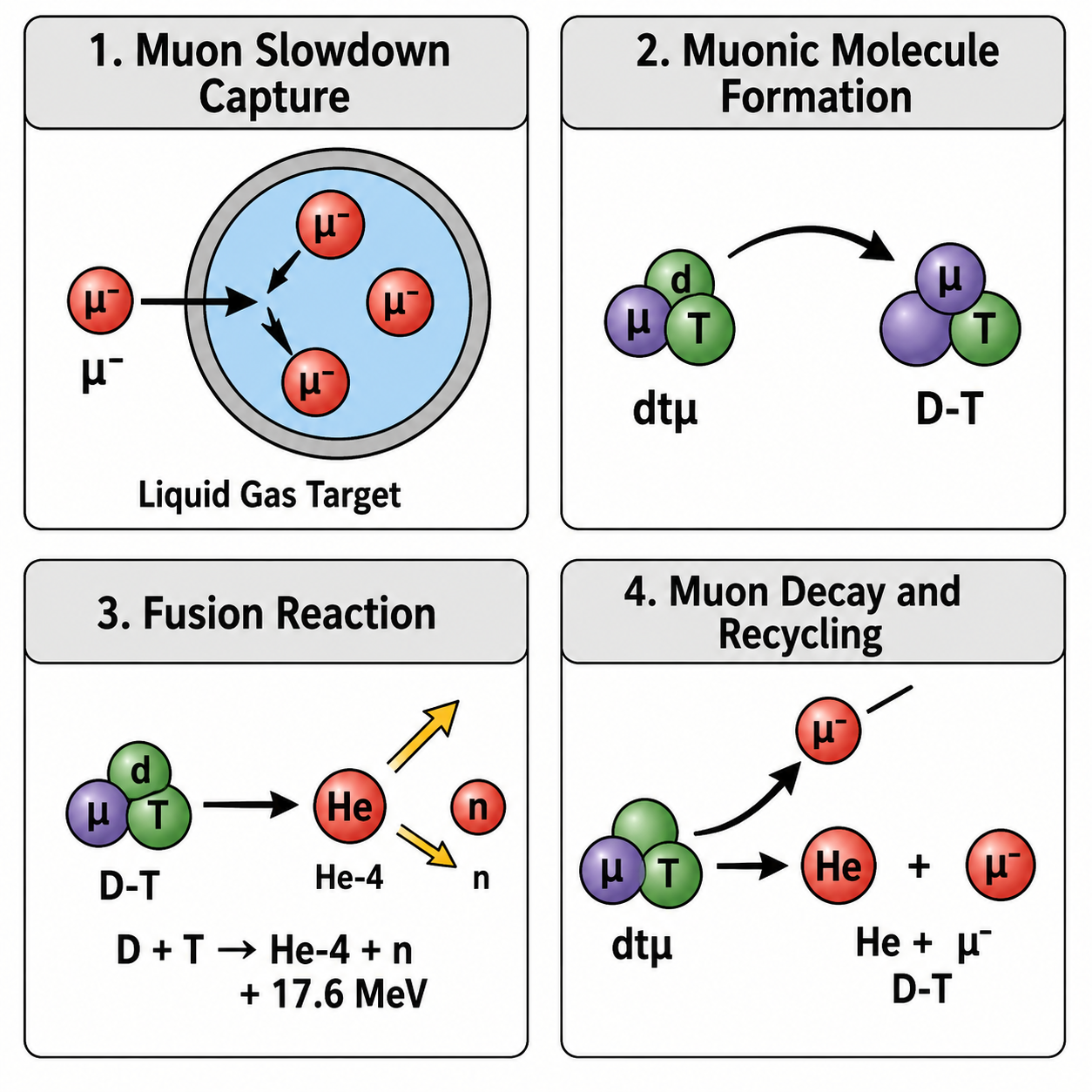}
		\caption{Four-step mechanism of muon-catalyzed D--T fusion.}
		\label{fig:cycle}
	\end{figure*}

	A complete muon-catalyzed fusion cycle can be decomposed into four key steps, referred to here as the four-step cycle; the corresponding schematic illustration is shown in Fig.~\ref{fig:cycle}.
	This process is a close interplay of atomic physics, molecular physics, and nuclear physics, and it also reveals the essential catalytic nature of the muon: ideally, the muon is not consumed after one fusion event.
	
	\subsection{Step I: Muonic-Atom Formation as the Starting Point of Orbital Compression}
	
	After a negative muon \(\mu^-\) is injected into a D--T fuel mixture, it is rapidly captured by a nucleus because its mass is much larger than that of the electron.
	A highly excited muonic atom, such as \(\mu t\) or \(\mu d\), is formed, while the original atomic electron is ejected through an Auger process \cite{Jones1986}:
	\begin{equation}
		\mu^- + T \rightarrow (\mu t)^+ + e^- .
	\end{equation}
	According to the Bohr model, the Bohr radius of a muonic atom is approximately \(1/207\) of that of a hydrogenic atom with an electron.
	This realizes the first stage of extreme compression at the atomic scale.
	Because the triton \(T\) has a larger mass number than the deuteron \(D\), its muon-capture cross section is larger, and the muon preferentially forms a \((\mu t)\) atom in a mixed fuel environment.
	
	\subsection{Step II: Muon Transfer and Optimization of the Catalytic Pathway}
	
	The electrically neutral \((\mu t)\) atom moves freely in the fuel gas.
	When it collides with a \(\mathrm{D}_{2}\) or D--T molecule, muon transfer can occur.
	In the simplified description used here, the muon is transferred from the triton to the deuteron, accompanied by the release of kinetic energy of approximately \(48~\mathrm{eV}\) \cite{Jones1986}:
	\begin{equation}
		(\mu t)+D \rightarrow (\mu d)+T+\Delta E,
		\qquad
		\Delta E \simeq 48~\mathrm{eV}.
	\end{equation}
	The transfer rate is of order \(10^{5}\)--\(10^{6}~\mathrm{s}^{-1}\), and it is regulated by fuel temperature and density.
	This process allows the muon to rapidly reorganize the level population in a \(\mathrm{D}_{2}\)--D--T mixture and thereby optimize the catalytic pathway automatically.
	Consequently, the cycle exhibits robustness against initial fuel-ratio variations.
	
	\subsection{Step III: Resonant \texorpdfstring{\(\dtmu\)}{dtmu} Molecular Formation as the Rate-Determining Step}
	
	As indicated schematically in Fig.~\ref{fig:cycle}, the most essential and subtle step of the catalytic cycle is the resonant formation of the \(\dtmu\) molecular ion:
	a \((\mu t)\) atom collides with a \(\mathrm{D}_{2}\) molecule and forms an excited muonic molecular ion \([(\dtmu)]^{+*}\).
	When the kinetic energy of the collision system matches the molecular binding energy, approximately \(0.66~\mathrm{eV}\), the \(\dtmu\) formation rate \(\lambda_{dt\mu}\) increases by orders of magnitude.
	This gives the process a strong temperature dependence and makes it the rate-determining step of the catalytic cycle \cite{Porcelli2001,Ackerbauer1999}:
	\begin{equation}
		(\mu t)+\mathrm{D}_{2}
		\rightarrow
		[(\dtmu)]^{+*}+D .
	\end{equation}
	
	In the muonic molecule, the deuteron and triton are bound by the muon at a separation of only about \(280~\mathrm{fm}\).
	The effective matter density, of order \(10^{8}~\mathrm{g\,cm^{-3}}\), is close to that in the interior of a white dwarf in a heuristic sense \cite{Breunlich1987,RafelskiMuller1985,Kamimura2023}.
	The internuclear distance is therefore strongly compressed.
	Experimentally, the resonant \(\dtmu\) molecular formation rate at \(800~\mathrm{K}\) can reach \((100\)--\(150)\times10^{8}~\mathrm{s}^{-1}\), about 50--75 times faster than that at room temperature.
	This is the central key for increasing the total catalytic-cycle rate.
	
	\subsection{Step IV: Fusion and Muon Release}
	
	At such a small internuclear separation, the deuteron and triton complete fusion through quantum tunneling on a very short time scale, approximately \(10^{-12}~\mathrm{s}\).
	The reaction releases \(17.6~\mathrm{MeV}\) of energy, producing a \(3.5~\mathrm{MeV}\) alpha particle, \({}^{4}\mathrm{He}^{2+}\), and a \(14.1~\mathrm{MeV}\) neutron:
	\begin{equation}
		(dt\mu)^+
		\rightarrow
		\alpha(3.5~\mathrm{MeV})
		+ n(14.1~\mathrm{MeV})
		+ \mu^- .
	\end{equation}
	Equivalently, the net nuclear reaction is
	\begin{equation}
		D+T
		\rightarrow
		{}^{4}\mathrm{He}+n+17.6~\mathrm{MeV}.
	\end{equation}
	
	After fusion, the \(\dtmu\) molecule breaks apart, and most muons are released and enter the next catalytic cycle.
	However, a critical loss mechanism exists:
	the muon can be captured by the fusion-produced alpha particle, forming a bound \((\alpha\mu)^+\) ion and permanently leaving the catalytic cycle.
	This is the alpha-sticking effect.
	Its probability \(\omega_s\) is the central limiting factor for the energy gain of \mucf{} \cite{Breunlich1987,RafelskiMuller1985,Kamimura2023}.
	As long as the muon is not stuck to the alpha particle, it remains available after the reaction and can catalyze another D--T fusion event.
	This highlights the role of the muon as a genuine quantum catalyst.
	
	\section{Energy-Balance Challenge and Breakthrough Toward \texorpdfstring{\(Q=1\)}{Q=1}}
	
	Although the physical principle of \mucf{} is clear, the central obstacle to engineering application is the energy balance.
	The key quantity for evaluating this balance is the energy-gain factor \(Q\), whose quantitative relation is given by the kinetic model of the catalytic cycle \cite{Iiyoshi2019,Yamashita2022}.
	
	\subsection{Quantitative Model for Catalysis Cycles per Muon and Energy Gain}
	
	During its lifetime, \(\tau_{\mu}\simeq2.197~\mu\mathrm{s}\), a single muon can catalyze a finite number of fusion events.
	Let \(\lambda_{\mu}=1/\tau_{\mu}\simeq4.55\times10^{5}~\mathrm{s}^{-1}\) be the muon decay rate, \(\lambda_c\) the catalytic-cycle rate, and \(\omega_s\) the alpha-sticking probability.
	The number of catalysis cycles per muon, \(X_{\mu}\), is modeled as
	\begin{equation}
		X_{\mu}
		=
		\frac{1}{\lambda_{\mu}/\lambda_c+\omega_s}.
		\label{eq:Xmu}
	\end{equation}
	Here, \(\lambda_{\mu}\) is the muon decay rate, \(\lambda_c\) is the effective catalytic-cycle rate, and \(\omega_s\) is the effective alpha-sticking probability.
	
	In present mainstream pion beam lines, such as PSI and J-PARC MUSE, the average effective energy cost for obtaining one negative muon is taken as
	\begin{equation}
		\langle E_{\mu}\rangle \simeq 5~\mathrm{GeV}.
	\end{equation}
	This value includes systematic losses from pion production, transport, decay, and muon collection \cite{YinChenKou2026NPR}.
	If one muon catalyzes \(X_{\mu}\) fusion events, the total fusion energy released is \(X_{\mu}\times17.6~\mathrm{MeV}\).
	The energy-gain factor is therefore
	\begin{equation}
		Q
		=
		\frac{X_{\mu}\times17.6~\mathrm{MeV}}{E_{\mu}} .
		\label{eq:Q}
	\end{equation}
	The landmark Los Alamos experiment measured approximately 150 catalysis cycles per muon, corresponding to \(\omega_s\simeq0.45\%\).
	Substitution into Eq.~\eqref{eq:Q} gives \(Q\simeq0.53\), still far below net energy production.
	The break-even condition \(Q=1\) requires at least approximately 284 D--T fusion events per muon.
	
	\subsection{Suppressing Alpha Sticking and Multi-Technology Synergy}
	
	There are two core routes to overcoming this bottleneck:
	reducing the alpha-sticking probability \(\omega_s\) and increasing the catalytic-cycle rate \(\lambda_c\).
	For the alpha-sticking problem, recent theoretical work has proposed a fuel-nucleus--muon dual-polarization control scheme \cite{YinChenKou2026NPR}.
	By polarizing both the D--T fuel nuclei and the muon beam, the formation of the \((\alpha\mu)^+\) bound state may be suppressed at the quantum level.
	Within the assumptions of the present kinetic model, full polarization alone is estimated to bring the following performance improvements:
	\begin{itemize}
		\item the alpha-sticking probability \(\omega_s\) decreases from approximately \(0.45\%\) to approximately \(0.34\%\);
		\item the catalytic-cycle rate \(\lambda_c\) increases by approximately \(30\%\)--\(50\%\);
		\item the number of catalysis cycles per muon \(X_{\mu}\) increases from 148 to approximately 193.
	\end{itemize}
	
	On this basis, we construct a four-dimensional synergistic kinetic model combining muon--fuel-nucleus dual polarization, high-density confinement, electric-field-assisted muon regeneration, and resonant enhancement \cite{YinChenKou2026NPR}.
	Under idealized assumptions, the model suggests the following possible trends:
	\begin{itemize}
		\item high-density confinement may further promote collisional stripping of \((\alpha\mu)^+\), potentially suppressing \(\omega_s\) below \(0.3\%\) \cite{Breunlich1987,RafelskiMuller1985,Kamimura2023};
		\item electric-field-assisted regeneration may actively recover part of the muon population captured by slow alpha particles;
		\item under multi-technology synergistic conditions, \(X_{\mu}\) may reach 300--350 and \(Q\) may exceed 1.1;
		\item according to our preliminary kinetic-model estimate, \(X_{\mu}\) may exceed 500 under engineering-limit optimized conditions, with \(Q\) potentially exceeding 2;
		\item if further advances in muon-lifetime control and decoherence suppression were realized, the number of catalysis cycles per muon could be further increased; however, this scenario remains a speculative upper-bound estimate.
	\end{itemize}
	These schemes should therefore be regarded as model-based projections rather than experimentally demonstrated performance benchmarks.
	Their experimental validation, especially the proposed reduction of \(\omega_s\) through dual polarization and assisted muon recovery, remains a central challenge.
	
	\begin{table*}[t]
		\caption{Calculated key parameters under different operating conditions.}
		\label{tab:conditions}
		\centering
		\scriptsize
		\setlength{\tabcolsep}{2.8pt}
		\renewcommand{\arraystretch}{1.25}
		
		\begin{tabular}{@{}lccccccc@{}}
			\toprule
			\makecell[c]{Parameter}
			&
			\makecell[c]{Symbol}
			&
			\makecell[c]{Unpolarized}
			&
			\makecell[c]{Fully\\polarized\\conservative}
			&
			\makecell[c]{Fully\\polarized\\optimistic}
			&
			\makecell[c]{Multi-technology\\synergy\\ultra-optimistic}
			&
			\makecell[c]{Engineering-limit\\optimized}
			&
			\makecell[c]{Ultimate\\theoretical\\condition}
			\\
			\midrule
			
			\makecell[l]{Alpha-sticking\\probability}
			&
			$\omega_s$
			&
			$0.0045$
			&
			$0.00342$
			&
			$0.00315$
			&
			$0.0018$--$0.0020$
			&
			$0.0008$--$0.0010$
			&
			$0.0006$
			\\
			\specialrule{0.25pt}{2pt}{2pt}
			
			\makecell[l]{Catalytic-cycle\\rate}
			&
			$\lambda_c~(\mathrm{s}^{-1})$
			&
			$2.0\times10^{8}$
			&
			$2.6\times10^{8}$
			&
			$3.0\times10^{8}$
			&
			$(3.2$--$3.3)\times10^{8}$
			&
			$(4.5$--$5.0)\times10^{8}$
			&
			$5.5\times10^{8}$
			\\
			\specialrule{0.25pt}{2pt}{2pt}
			
			\makecell[l]{Muon decay\\rate}
			&
			$\lambda_\mu~(\mathrm{s}^{-1})$
			&
			$4.55\times10^{5}$
			&
			$4.55\times10^{5}$
			&
			$4.55\times10^{5}$
			&
			$4.55\times10^{5}$
			&
			$4.55\times10^{5}$
			&
			$3.0\times10^{5}$
			\\
			\specialrule{0.25pt}{2pt}{2pt}
			
			\makecell[l]{Catalysis cycles\\per muon}
			&
			$X_\mu$
			&
			$148$
			&
			$193$
			&
			$292$
			&
			$300$--$350$
			&
			$500$--$600$
			&
			$873$
			\\
			\specialrule{0.25pt}{2pt}{2pt}
			
			\makecell[l]{Energy\\gain}
			&
			$Q$
			&
			$0.52$
			&
			$0.68$
			&
			$1.03$
			&
			$1.06$--$1.23$
			&
			$1.76$--$2.11$
			&
			$3.07$
			\\
			
			\bottomrule
		\end{tabular}
		
		\vspace{0.6em}
		
		\begin{minipage}{0.96\textwidth}
			\footnotesize
			\textit{Note.}
			The multi-technology-synergy column assumes the simultaneous action of dual polarization, high-density confinement, electric-field-assisted muon recovery, and resonant enhancement.
			The last two columns, ``engineering-limit optimized'' and ``ultimate theoretical condition,'' are extrapolative predictions based on the present kinetic model.
			Their technical realization would depend on frontier directions such as muon-lifetime control and strong decoherence suppression, which have not yet been experimentally demonstrated.
			Economic assessment should therefore be based primarily on the range from the unpolarized case to the multi-technology-synergy case, where experimental support or a clearer technical route exists.
		\end{minipage}
		
	\end{table*}
	
	The next experimental directions include:
	\begin{itemize}
		\item constructing a polarized D--T target experimental station based on the Huizhou CiADS high-intensity muon source EMuS, with a post-2028 flux expected to reach at least \(10^{8}~\mu^{-}/\mathrm{s}\);
		\item precisely measuring the dependence of \(\omega_s\) on the nuclear polarization degrees \(P_D\) and \(P_T\) and the muon polarization degree \(P_{\mu}\), in order to test whether \(\omega_s\) can be reduced from \(0.45\%\) to below \(0.3\%\);
		\item studying depolarization during muonic-atom formation, muon transfer, and resonant molecular formation, and extending the polarization retention time using superconducting magnetic fields and laser pumping;
		\item collaborating with international muon facilities such as J-PARC MUSE and PSI to share polarized-target technology and data.
	\end{itemize}
	
	\section{Technical Schemes and Future Outlook}
	
	\subsection{In-Flight Muon-Catalyzed Fusion and Heavy-Ion-Driven Fusion}
	
	To bypass the bottleneck associated with \(\dtmu\) molecular formation in the conventional catalytic cycle, Japanese researchers proposed the concept of in-flight muon-catalyzed fusion (IF\mucf).
	In this scheme, high-kinetic-energy \((\mu t)\) atoms, typically in the range \(1\)--\(10~\mathrm{keV}\), collide directly with deuterium nuclei.
	Fusion can then be triggered without first forming a bound muonic molecule \cite{Iiyoshi2019,Liu2022IFMCF}.
	This route not only partially avoids the premises of the alpha-sticking effect but also broadens the applicable operating conditions of \mucf{}.
	
	In parallel, the team led by Wenlong Zhan has proposed a heavy-ion-driven magneto-inertial fusion (HIMIF) route \cite{Zhan2024HIMIF}.
	In this scheme, intense heavy-ion accelerator facilities such as HIAF generate high-energy heavy-ion beams that strike high-\(Z\) targets to produce pions.
	The pions then decay into muons, while a magnetized high-energy-density plasma is formed.
	The scheme attempts to integrate inertial confinement, magnetic confinement, and muon catalysis into a unified framework.
	
	\subsection{A Hybrid Reactor Concept: \texorpdfstring{\mucf}{muCF}-FBR}
	
	Based on the compactness of \mucf{} devices, the high neutron energy of \(14.1~\mathrm{MeV}\), and the concentrated neutron flux, this paper proposes a fusion--fission fuel-breeding hybrid reactor concept:
	\mucf-FBR (Muon-Catalyzed Fusion--Fission Fuel Breeding Hybrid Reactor).
	Its core idea is to realize decoupled fusion--fission operation and to build a neutron factory dedicated to nuclear-fuel breeding \cite{YinChen2026Book}.
	At the present stage, \mucf-FBR should be regarded as a conceptual reactor design; detailed neutron-transport, burnup, thermal-hydraulic, and materials-damage calculations remain to be performed.
	Below, the concept is analyzed from three aspects: physical mechanism, engineering implementation, and multidimensional value.
	
	\subsubsection{Physical Mechanism and Decoupled Design}
	
	Conventional fusion--fission hybrid reactors usually couple fusion and fission tightly in both space and time, which leads to extreme engineering complexity.
	The central innovation of \mucf-FBR is to decouple the two processes completely.
	
	On the fusion side, namely the neutron source, D--T fusion reactions occur continuously in the \mucf{} reaction vessel and produce \(14.1~\mathrm{MeV}\) fast neutrons.
	The task of this subsystem is to serve as a stable and controllable intense neutron source.
	Although fission heat may still be recovered, the extraction of \({}^{239}\mathrm{Pu}\) is the main objective.
	
	On the breeding side, namely the fuel factory, a breeding layer made of natural uranium or depleted uranium is arranged around the fusion vessel.
	Its dominant component is the fertile isotope \({}^{238}\mathrm{U}\).
	After fast neutrons enter the breeding layer, the reaction chain
	\begin{equation}
		{}^{238}\mathrm{U}(n,\gamma){}^{239}\mathrm{U}
		\rightarrow
		{}^{239}\mathrm{Np}
		\rightarrow
		{}^{239}\mathrm{Pu}
	\end{equation}
	converts \({}^{238}\mathrm{U}\) efficiently into fissile \({}^{239}\mathrm{Pu}\).
	
	In the decoupled operating and offline-extraction mode, the fusion reactor operates for a certain irradiation period and is then shut down.
	The \({}^{239}\mathrm{Pu}\)-rich breeding layer is removed for chemical separation and fuel-element fabrication, and the resulting fuel is supplied to conventional fission reactors.
	This ``fusion makes fuel, fission uses fuel'' operating mode is the essence of \mucf-FBR.
	
	\subsubsection{Engineering Feasibility and Advantages}
	
	This decoupled design converts the intrinsic physical characteristics of \mucf-FBR into substantial engineering advantages.
	
	First, it helps avoid the first-wall problem.
	The first wall of a fusion reactor must withstand intense high-energy neutron irradiation, which leads to material transmutation and embrittlement.
	In \mucf-FBR, the breeding layer is designed as a periodically replaceable sacrificial component.
	This greatly reduces the material requirements for the fusion vessel itself and extends the lifetime of the core device.
	
	Second, the engineering complexity is simplified.
	Because fission reactions occur in a different spatial and temporal domain, the \mucf{} device does not need to address compatibility with fission fuel, coolant compatibility, corrosion by fission products, and related issues.
	The difficulty of engineering design is therefore greatly reduced, allowing a more compact and modular reactor architecture.
	
	Third, the utilization of nuclear fuel is improved and the fuel-supply problem is alleviated.
	Only \(0.7\%\) of natural uranium is the fissile isotope \({}^{235}\mathrm{U}\).
	The \mucf-FBR concept can convert the abundant \({}^{238}\mathrm{U}\), which accounts for \(99.3\%\) of natural uranium, into \({}^{239}\mathrm{Pu}\).
	In principle, this can increase uranium-resource utilization by more than two orders of magnitude, fundamentally changing the nuclear-fuel supply structure and providing fuel security for sustainable global nuclear energy \cite{YinChenKou2026NPR}.
	
	\subsubsection{Multidimensional Strategic Value}
	
	The value of \mucf-FBR goes beyond energy production alone and has multidimensional strategic significance, as summarized in Table~\ref{tab:value}.
	
	\begin{table*}[t]
		\caption{Strategic value of the \mucf-FBR concept.}
		\label{tab:value}
		\centering
		\footnotesize
		\setlength{\tabcolsep}{3.5pt}
		\renewcommand{\arraystretch}{1.25}
		
		\begin{tabular}{@{}llll@{}}
			\toprule
			\makecell[l]{Value\\dimension}
			&
			\makecell[l]{Core\\function}
			&
			\makecell[l]{Expected\\effect}
			&
			\makecell[l]{Technical or\\design key}
			\\
			\midrule
			
			\makecell[l]{Energy\\security}
			&
			\makecell[l]{Fuel breeding\\and supply\\assurance}
			&
			\makecell[l]{Provides an autonomous\\nuclear-energy route\\for uranium-poor\\countries}
			&
			\makecell[l]{${}^{238}\mathrm{U}\rightarrow{}^{239}\mathrm{Pu}$\\conversion efficiency\\$>90\%$}
			\\
			\midrule
			\makecell[l]{Nuclear-waste\\transmutation}
			&
			\makecell[l]{Treatment of\\long-lived\\fission products}
			&
			\makecell[l]{Reduction of\\high-level waste\\volume; transmutation\\rate $>90\%$}
			&
			\makecell[l]{14.1-MeV fast-neutron\\flux $>10^{15}$\\$\mathrm{n\,cm^{-2}\,s^{-1}}$}
			\\
			\midrule
			\makecell[l]{Nonproliferation\\potential}
			&
			\makecell[l]{Control of\\plutonium isotopic\\composition}
			&
			\makecell[l]{Potential increase in\\proliferation resistance\\through control of\\plutonium isotopic\\composition}
			&
			\makecell[l]{Precise control\\of uranium enrichment\\in the breeding layer}
			\\
			\midrule
			\makecell[l]{Technology\\pull}
			&
			\makecell[l]{Frontier-technology\\integration\\platform}
			&
			\makecell[l]{Drives development of\\radiation-resistant\\first-wall and breeding-layer\\materials, high-flux\\muon sources, and\\related key technologies}
			&
			\makecell[l]{Multidisciplinary\\research and\\development demand}
			\\
			
			\bottomrule
		\end{tabular}
		
		\vspace{0.6em}
		
		\begin{minipage}{0.96\textwidth}
			\footnotesize
			\textit{Note.}
			The nonproliferation assessment is qualitative at this stage.
			A quantitative evaluation of plutonium isotopic vectors should be performed in future burnup and safeguards analyses; IAEA INFCIRC/549 is cited here only as a general reference for plutonium-management policy \cite{IAEA_INFCIRC549}.
			The material-irradiation parameters are extrapolated from ITER blanket-design experience.
		\end{minipage}
		
	\end{table*}
	
	\section{Conclusion}
	
	Muon-catalyzed nuclear fusion opens a subatomic-scale fusion pathway distinct from the high-temperature plasma-fusion paradigm.
	Its physical essence is not simply to lower the temperature.
	Rather, through muon-mass-induced quantum orbital compression, the D--T internuclear distance is actively driven into the strong-interaction-dominated region, enabling nuclear fusion under thermodynamic room-temperature conditions.
	This is the strict physical meaning of the ``miniature Sun.''
	
	Although \mucf{} was once regarded as an elegant but impractical scientific curiosity because of the short muon lifetime and the high energy cost of muon production, recent interdisciplinary progress is systematically pushing the balance toward \(Q>1\).
	Key advances include alpha-sticking suppression, resonance optimization, and high-intensity muon-source development.
	The quantitative analysis in this work indicates that, through multi-technology synergy, the number of catalysis cycles per muon may increase stepwise to approximately 300, 500, and even 800, thereby breaking the energy-balance bottleneck.
	
	On this basis, the \mucf-FBR concept proposed here positions \mucf{} as a fusion neutron source.
	Through a decoupled fusion--fission mode, it can in principle breed nuclear fuel efficiently.
	The key engineering value of this decoupled design is that the \mucf{} subsystem functions primarily as a controllable neutron source, while fuel breeding and fuel utilization are handled in a separate blanket and downstream fission-fuel cycle.
	This strategy avoids the stringent physical requirements of direct \mucf{} power generation and transforms the present physical reality of \mucf{} into substantial engineering and strategic value.
	The four-step cycle embodied in \mucf{} may therefore first play its strongest role in igniting the fuel cycle of future advanced nuclear-energy systems.
	
	\bibliographystyle{apsrev4-2}
	\bibliography{refs}
	
\end{document}